\documentclass[11pt,a4paper]{article}
\pdfoutput=1
\usepackage{jheppub}
\usepackage{amsmath,amssymb,latexsym,revsymb,color,hyperref}

\usepackage{amsthm}

\def\<{\langle}
\def\>{\rangle}

\newcommand{\be}{\begin{eqnarray} \begin{aligned}}
\newcommand{\ee}{\end{aligned} \end{eqnarray} }
\newcommand{\benn}{\begin{eqnarray*} \begin{aligned}}
\newcommand{\eenn}{\end{aligned} \end{eqnarray*} }

\newcommand{\ben}{\begin{eqnarray} \begin{aligned}}
\newcommand{\een}{\end{aligned} \end{eqnarray} }

\newcommand{\bc}{\begin{center}}
\newcommand{\ec}{\end{center}}

\newcommand{\id}{\mathbb{I}}

\newcommand{\tr}{\mathop{\mathsf{tr}}\nolimits}

\newcommand{\beq}{\begin{eqnarray} \begin{aligned}}
\newcommand{\eeq}{\end{aligned} \end{eqnarray} }
\newcommand{\bea}{\begin{array}}
\newcommand{\eea}{\end{array}}

\newcommand{\bee}{\begin{enumerate}}
\newcommand{\eee}{\end{enumerate}}
\newcommand{\bei}{\begin{itemize}}
\newcommand{\eei}{\end{itemize}}

\usepackage{amsfonts}

\def\id{\mathbb{I}}

\def\01{\{0,1\}}

\newcommand{\ket}[1]{|#1\rangle}
\newcommand{\bra}[1]{\langle#1|}

\newcommand{\proj}[1]{|#1\rangle\langle#1|}

\def\<{\langle}
\def\>{\rangle}

\newtheorem*{rep@theorem}{\rep@title}
\newcommand{\newreptheorem}[2]{%
\newenvironment{rep#1}[1]{%
 \def\rep@title{#2 \ref{##1} (restatement)}%
 \begin{rep@theorem}}%
 {\end{rep@theorem}}}
\makeatother

\newreptheorem{thm}{Theorem}
\newreptheorem{lem}{Lemma}

\begin{document}
\title{Firewalls and flat mirrors: An alternative to the AMPS experiment which evades the Harlow-Hayden obstacle}
\author[a]{Jonathan Oppenheim,}
\author[b]{Bill Unruh}
\affiliation[a]{University College of London,\\ Department of Physics \& Astronomy, London, WC1E 6BT }
\affiliation[b]{University of British Columbia,\\Department of Physics \& Astronomy, Vancouver}
\abstract{If quantum gravity does not lead to a breakdown of predictability, then Almheiri-Marolf-Polchinski-Sully (AMPS) have argued that an observer falling into a black hole can perform an experiment which verifies a violation of entanglement monogamy -- that late time Hawking radiation is maximally entangled with early time Hawking radiation and also with in-falling radiation -- something impossible in quantum field theory. However, as pointed out by Hayden and Harlow, this experiment is infeasible, as the time required to perform the experiment is almost certainly longer than the lifetime of the black hole. Here we propose an alternative firewall experiment which could actually be performed within the black hole's lifetime. The alternative experiment involves forming an entangled black hole in which the unscrambling of information is precomputed on a quantum memory prior to the creation of the black hole and without acting on the matter which forms the black hole or emerges from it. This would allow an observer near a black hole to signal faster than light. As another application of our precomputing strategy, we show how one can produce entangled black holes which acts like  ``flat mirrors'', in the sense that information comes out almost instantly (as in the Hayden-Preskill scenario), but also emerge unscrambled, so that it can actually be observed instantly as well. 
Finally, we prove that a black hole in thermal equilibrium with its own radiation, is uncorrelated with this radiation.}
\date{\today}\footnote{the initial findings of this paper were originally communicated in January 2012, and a presentation of the results is available at http://online.kitp.ucsb.edu/online/fuzzorfire-m13/oppenheim/ (KITP, August 22, 2013)} 
\maketitle

\section{Is the AMPS experiment feasible?}

\begin{figure}
\includegraphics[width=10cm]{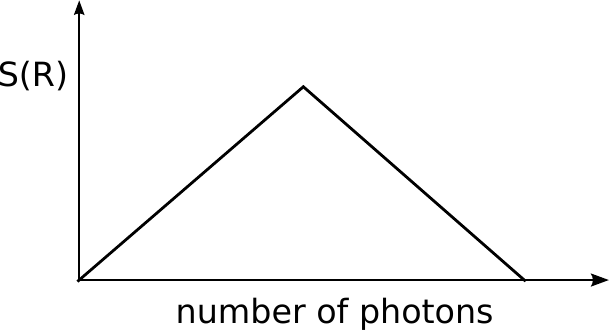}
	\caption{	
	If a black hole is initially a pure state, and we look at the Von Neumann entropy $S(R_m):=-\tr\rho_R\log\rho_R$ of all the particles $R_m$ in state $\rho_R$ which have radiated from the black hole as function of the number of radiated particles $m$, then we expect to get something like the diagram pictured above\cite{page-unitary-evap}.  The Page time $t_{Page}$ is defined as the apex of the triangle after which the black hole radiation must be entangled with the early time radiation. This must occur because if information is preserved, then a pure state must evolve to a pure state, and at the end of the evaporation process, the state of the radiation must return to being pure with entropy zero. From the point of view of the outside observer, all points on the line after the Page time correspond to points in which the late radiation is uncorrelated with the black hole.  This is because the line represents the point where each particle is reducing the entropy of the previously emitted radiation by its dimension. I.e. if at time $t\geq t_{Page}$ the black hole emits some particles $B$, then the line represents the point $S(R_mB)=S(R_m)-S(B)$. Since the purification of the state on $R_mB$ is the black hole $H$, we have $S(R_mB)=S(H)$, $S(R_t)=S(BH)$ and consequently $I(B:H):=S(B)+S(H)-S(BH)=0$, and thus $B$ is uncorrelated with the black hole.  
	}
\label{fig:page}
\end{figure}

The AMPS experiment~\cite{almheiri2013black} (c.f. also \cite{braunstein2009entangled}) sharpens the black hole information problem~\cite{hawking-bhinfoloss,hawking-unpredictability,preskill-infoloss-note}. If black holes preserve information, then this seems to imply that either black holes have a firewall just below the horizon, or the density matrix of fields on the black hole space time evolves to something which is not a density matrix, since there is a violation of a principle known as the monogamy of entanglement\cite{coffman2000distributed,bennet-monogomy,koashi2004monogamy} -- something impossible for a quantum density matrix. Monogomy of entanglement is the property that if two systems are maximally entangled, then neither of them can be correlated with anything else. In the case of black holes, outgoing Hawking radiation is maximally entangled with infalling Hawking radiation (heuristically, one can think of entangled particle-anti-particle pairs, one which escapes the black hole and one which falls down the black hole). However, if black holes preserve information, then late time outgoing Hawking radiation must also be entangled with earlier outgoing Hawking radiation, in contradiction with it also being entangled with the infalling radiation. The reason information preservation requires late time Hawking radiation to be entangled with early time radiation, is that we can imagine forming a black hole in an initially pure quantum state, and after it has radiated half its photons thermally, the remaining photons must be entangled with the previous one, in order to preserve the purity of the full state (see Figure \ref{fig:page}).

AMPS argued that one could witness this violation of monogamy of entanglement, by having an observer Alice, who collects the early time Hawking radiation $R$. She then jumps into the near horizon region of the black hole (now commonly referred to as 'the zone'), to verify that her radiation is maximally entangled with the photons there (which we denote by $B$). This contradicts the fact that $B$ needs to be maximally entangled with ${\hat B}$, the infalling Hawking radiation. Alice can verify that $B$ and $\hat B$ are maximally entangled, because if they weren't, she would witness a firewall, since the state in which $B$ and $\hat B$ are uncorrelated (ia product state), has extremely high energies just inside the horizon. Previous proposals to circumvent the black hole information problem, such as black hole complementarity~\cite{tHooft-bhcompl,tHooft-bhcompl-string,susskind-bhcompl} do not allow one to escape the AMPS experiment, which in our view gives a strong motivation to further explore theories in which information is fundamentally destroyed~\cite{bps,unruh-wald-onbps,OR-intrinsic,unruh2012decoherence}.

The AMPS experiment has generated a lot of interesting discussion~\cite{marolf2013gauge,giddings2012nonviolent,susskind2012singularities,papadodimas2012infalling,bousso2013complementarity,jacobson2012boundary,banks2013no,almheiri2013apologia,maldacena2013cool,shenker2013black,mathur2013flaw,van2013evaporating,larjo2013black,lloyd2013unitarity}, but it has recently
been observed that the AMPS experiment as originally conceived, cannot be carried out~\cite{harlow2013quantum,susskind2013black}. This is because Alice needs to perform a quantum computation on the Hawking radiation in order to decode entanglement between the zone and the early Hawking radiation, and the time that it would take to perform the quantum computation is longer than the time it takes for the black hole to evaporate. By the time Alice is ready to jump into the black hole to verify the violation of entanglement monogamy, it is too late, and the black hole has gone. Now, even if it were impossible to witness a violation of monogamy of entanglement, we still believe the mere fact of the violation would be problematic enough. This is because it would imply that the density matrix has evolved into something which is not a density matrix (in fact, one can show that it would typically have negative eigenvalues in order to violate monogomy). It would be very difficult to construct a theory in which density matrices didn't evolve into density matrices, whether one could verify this via a physically implementable measurement or not. Nonetheless, it is an interesting question whether one could witness a violation of the monogamy of entanglement if black holes preserve information. In the present paper we propose an experiment to do so. It has also been argued that if black holes preserve information, then it 
may be impossible to verify this, since it would require clocks which are so accurate, they'd form black holes\cite{unruh-newton}. Our proposed experiment circumvents that obstacle as well.

In the remainder of this section, we briefly review the Harlow-Hayden obstruction to the original AMPS experiment. In Section \ref{sec:alternative} we propose the first of three alternative gedanken experiments using a maximally entangled black hole.  The key point is that rather than create an entangled black hole by letting it radiate until the black hole is maximally entangled with its own radiation, and having to perform a decoding operation on the emitted radiation, one instead creates the black hole such that it is initially entangled with a quantum memory and one can perform the computation on the memory even before the black hole is formed. How to construct such a maximally entangled black hole is discussed in Section\ref{sec:entangledbh}. In Sections \ref{sec:box} and \ref{sec:robust} we discuss two other gedanken experiments. The one in Section \ref{sec:box} works for more exotic models of black hole evaporation, and the final one is more robust against imperfections or disturbances in the set-up (for example, errors made while forming the black hole, in which an external particle inadvertently falls into the black hole). Finally, in Section \ref{sec:chsh} we show how one can use the entangled black hole to signal faster than light. In Section \ref{sec:product} of the Appendix, we prove that if a black hole is in thermal equilibrium with its own Hawking radiation, then the correlations between the black hole and the radiation must be vanishingly small.  Appendix \ref{sec:general} contains some more technical aspects of our result, in particular we provide criteria for when a black hole is decoupled from its own Hawking radiation which holds for all possible models of black hole evaporation. For the remainder of the paper, we assume that black hole evaporation is unitary, in order to show that this will lead to a contradiction, either with monogamy of entanglement, or with any of the postulates of AMPS~\cite{almheiri2013black}.

Returning now to the AMPS experiment, we denote the black hole by $H$ and the photons which have radiated away at early time, we denote by $R$. We denote by $B$, the late time photons which have evaporated into the zone. We are interested in the state of these three systems some time after the black hole has radiated away more than half its photons. I.e. we want to wait until the black hole as evaporated long enough, such that the late time Hawking radiation $R$ is maximally entangled with the black hole $H$ and the photons in the zone $B$. This is known as the Page time~\cite{page-unitary-evap} (see Figure \ref{fig:page}), and the total state of the system is given by
\begin{align}
\ket{\Psi}_{AMPS}=\frac{1}{\sqrt{|B||H|}}
=\sum_{h,b}\ket{b}_B\ket{h}_H U_R\ket{bh0}_{R}
\label{eq:hh}
\end{align}
where we include an additional register in the $\ket{0}$ state, because the states $U_R\ket{bh0}_{R}$ might lie in a larger Hilbert space than the $\ket{bh}$ states of the black hole and zone. The basis $\ket{bh}$ can be written as pairs of entangled particles 
\begin{align}
\ket{bh}=\ket{b_1b_2...b_kh_1h_2...h_{n-k}}
\end{align}

If Alice can apply  $U^\dagger_R$ to the  Hawking radiation $R$ then she can succeed in distilling a pure entangled state 
between some small subset $R_B$ of the early time Hawking radiation, and subsystem $B$.
\begin{align}
\id_{HB}\otimes U^\dagger_R\ket{\Psi}_{AMPS}= \frac{1}{\sqrt{|B|}}\sum_{b}\ket{b}_B\ket{b}_{R_B}\otimes\frac{1}{\sqrt{|H|}}\sum_{h}\ket{h}_H\ket{h}_{R_H}
\end{align}
In fact, she will have pairs of entangled bits on systems $B_1,B_2,...B_k$ each individually entangled with  $R_{B_1},R_{B_2},...R_{B_k}$. Here $B$ can be considered to be the near horizon radiation in the zone. However, as observed by Harlow and Hayden, the unitary $U_R$ is likely to be very complicated from a computational point of view, and it is almost certain that implementing it will take longer than the lifetime of the black hole. Alice will not have enough time before she must dive into the black hole to observe a violation of entanglement monogamy. The AMPS experiment does not appear feasible, even theoretically.

\section{An alternative firewall experiment using a quantum memory entangled with a black hole}
\label{sec:alternative}

The problem with the original AMPS experiment, is that Alice has to perform her decoding operations on the Hawking radiation, and she has to do so before the black hole evaporates. However, we will see that in order to decode pure entanglement, she does not need to perform her decoding operation on the radiation, but can instead act on a quantum memory, and she can do so even before the black hole is created. We will first present a simple protocol which allows her to verify a violation of monogamy of entanglement. 
It works for most models of black hole evaporation, but one can imagine models of black hole evaporation for which it doesn't work. We will thus also present two other protocols in Section \ref{sec:box} and \ref{sec:robust}, which are slightly more complicated, but which can work for more exotic black hole evaporation models and are more robust against errors.

Let us first consider a simplified situation -- we imagine that we have a black hole (BH), entangled with a quantum memory $M$ (we discuss forming such an entangled black hole in Section \ref{sec:entangledbh} ). Our memory $M$ is composed of two quantum registers $M_H$, and $M_B$  and we start with the initial state
\begin{align}
\ket{\Psi(0)}=\frac{1}{\sqrt{|B||H|}}
\sum_{h,b}\ket{b}_{M_B}\ket{h}_{M_H} W_{BH} \ket{bh}_{BH}
\label{eq:initial}
\end{align}
where here, we take $W_{BH}\ket{bh}_{BH}$ to initially be the state of a black hole with $W_{BH}$ being the unitary of the formation
process.  Here, $H$ and $B$ are taken initially be inside the black hole.  
We next imagine that some particles evaporate from the black hole. We can take this process to be a unitary $U_{BH}$ which acts on the interior degrees of freedom of the black hole $BH$ followed by the ejection of system $B$ into the zone.  I.e. if we describe the process as a map on the interior of the black hole, then it is taken to be
\begin{align}
{\cal T}(\rho_{BH})=\tr_B \left( U_{BH}\rho_{BH}U^\dagger_{BH}\right)
\label{eq:page-map}
\end{align}
This is the type of black hole models generally used (e.g. \cite{page-unitary-evap,bhlock,HaydenPreskill,sekino2008fast,giddings2013quantum}) and we will consider more exotic ones later.  In the case where the black hole is maximally entropic (i.e. the entropy of the black hole is maximal for the mass of the black hole), we can describe the state of the black hole as $\id_{BH}/|B||H|$ i.e. the black hole is maximally mixed on a system of dimension $|B||H|$. In such a case, we have 
\begin{align}
||\rho_{BH}-\rho_{H}\otimes\rho_{B}||\leq \epsilon
\label{eq:productB}
\end{align}
because the maximally mixed state is invariant under a unitary transformation, and every part of it is uncorrelated with every other part. 
Equation \eqref{eq:productB} provides a necessary and sufficient condition for decoding entanglement between the zone photons and the memory, by acting only on the memory. To see this, we observe, following\cite{how-merge,how-merge2} that one possible purification
\footnote{A purification of $\rho_X$ is any pure state $\psi_{XY}$, such that $\rho_X=\tr_Y\psi_{XY}$.}
 of $\rho_{H}\otimes\rho_{B}$ is $\psi_{HM_{\overline{H}}}\otimes
\psi_{BM_{\overline{B}}}$,  where $M_{\overline{H}}$, $M_{\overline{B}}$ are subsystems of the memory. Then, since all purifications are unique up to a unitary
on the purifying system, this implies that there exists a unitary $V_{M}$ such that
\begin{align}
V_{M}\otimes U_{BH}\ket{\Psi(0)}_{BHM_{BH}}=\ket{\psi}_{HM_{\overline{H}}}\otimes\ket{\psi}_{BM_{\overline{B}}}
\end{align}
where $V_{M}$ is the decoding map that Alice needs to apply to the quantum memory in order to decode pure state
entanglement between the zone photons and the memory $M_B$.  Here, $V_{M}$ acts on systems $M_H$,$M_B$, and we denote the output systems
by $M_{\overline{B}}$ and $M_{\overline{H}}$, adopting the convention of using a bar to denote the output space. Thus after Alice's application of $V_{M}$, the total state 
consists of decoded entanglement between her quantum memory and the zone as required: 
\begin{align}
\ket{\Psi'}=\frac{1}{\sqrt{|B|}}
\sum_{h,b}\ket{b}_{M_{\overline B}}\ket{b}_B\otimes \frac{1}{\sqrt{|H|}}\ket{h}_{M_{\overline H}} \ket{h}_{H}
\label{eq:decoded}
\end{align}
Equation \eqref{eq:decoded}, is guaranteed by Uhlmann's theorem\cite{uhlmann1976transition}, which relates the L1 norm used in Equation \eqref{eq:productB} to the distance between the purifications of the states. That Equation \eqref{eq:productB} is a necessary condition, follows from the fact that if $\rho_B$ is correlated with $H$, then it cannot be pure and entangled with another system, and we will examine this quantitatively in Section \ref{sec:robust},
to show that a violation of monogamy of entanglement, need not require decoding pure state entanglement between the zone and the memory.

Now, the computational complexity of applying $V_{M}$ is not likely as great as for the $U^\dagger_R$ that Alice needed to apply to decode entangled pairs starting from
the state of Equation \eqref{eq:hh}, but it may still take her a long time to implement it. However, the key difference here, compared with the scenario considered by AMPS and Harlow-Hayden, is that the unitary $V_{M}$ she has to apply does not act on the radiation, but rather on the quantum memory. She can thus act 
on her quantum memory, even before the black hole is created. Thus, she sets up a situation where the creation of the black hole, and the subsequent evaporation of the black hole, result in pre-decoded radiation being emitted from the black hole, already perfectly correlated with individual qubits in the quantum memory. Having performed this precomputation on her memory, the process of black hole creation and evaporation performs the decoding for her.

Detailing the protocol more explicitly (see Equation \ref{eq:protocol} for the state after each of the steps): (i)
Alice begins by preparing a maximally entangled state shared between two quantum memories $M$ and $N$ with registers denoted by subscripts $B$ and $H$.
(ii) She then acts the unitary $V_{M}$  on her memory $M$. (iii) Next, she copies the state of the $N$ memory onto physical matter $BH$ and then she collapses this system into a black hole. The dynamics of this we denote by $W_{BH}$, and the protocol for creating this system is described in Section \ref{sec:entangledbh}. (iv) Next, she waits for the black hole to evaporate some photons into the zone, which is equivalent to letting $U_{BH}$ act on the black hole, followed by ejection of the system $B$ into the zone. The resulting state of the memory and black hole system is that of a pure entangled state with decoded entanglement between the zone and memory register $M_B$. The state at each step is given below, with the subscript on the state denoting each step of the protocol
\begin{align}
\ket{\Psi}_i&=\frac{1}{\sqrt{|B||H|}}
\sum_{h,b}\ket{b}_{M_B}\ket{h}_{M_H}\ket{b}_{N_B}\ket{h}_{N_H}\nonumber\\
\ket{\Psi}_{ii}&=\frac{1}{\sqrt{|B||H|}}
\sum_{h,b} V_M\ket{b}_{M_B}\ket{h}_{M_H}\ \ket{b}_{N_B}\ket{h}_{N_H}\nonumber\\
\ket{\Psi}_{iii}&=\frac{1}{\sqrt{|B||H|}}
\sum_{h,b} V_M\ket{b}_{M_B}\ket{h}_{M_H}W_{BH} \ket{bh}_{BH}\nonumber\\
\ket{\Psi}_{iv}&=\frac{1}{\sqrt{|B||H||E|}}
\sum_{h,b} V_M\ket{b}_{M_B}\ket{h}_{M_H}\ket{e}_{M_E}U_{BH}W_{BH} \ket{bh}_{BH} \nonumber\\
\label{eq:protocol}
\end{align} 
the final state $\ket{\Psi}_{iv}$ is equal to the decoded state of Equation \eqref{eq:decoded} as required.

Now, it may not be possible to create the initial state of Equation \eqref{eq:initial}. This is because in all likelihood, during the process of growing the black hole, the black hole will be evaporating at the same time. However, this is not an obstacle. Denoting these escaped particles by $E$ and an additional quantum memory by $M_E$,
we will instead have the initial state
\begin{align}
\ket{\Psi(0)}=\frac{1}{\sqrt{|B||H||E|}}
\sum_{h,b}\ket{b}_{M_B}\ket{h}_{M_H}\ket{e}_{M_E} W_{BHE} \ket{bhe}_{BHE}
\label{eq:initialE}
\end{align}
where we take $W_{BHE}\ket{bhe}_{BHE}$ to initially be the state of a black hole and escaped radiation with $W_{BHE}$ being the unitary of the formation
process.  Now, all we require after evaporation of the system $B$ into the zone is
\begin{align}
||\rho_{BHE}-\rho_{HE}\otimes\rho_{B}||\leq \epsilon
\label{eq:productBE}
\end{align}

This will be the case for an evaporation model like Equation \eqref{eq:page-map} where a unitary is applied to the internal degrees of freedom and then a system is ejected into the zone, provided that the interactions between $B$ and $E$ are small. I.e. in such a model $E$ just represents previous particles which have escaped the black hole, and they will thus be in a product state with $B$ and $H$. Essentially, we may just treat the system $E$ as being with $H$ for the purposes of Alice's decoding. This is described in more detail in the following section. It is also worth noting
that Equation \eqref{eq:productBE} correctly describes a black hole in thermal equilibrium as shown in Section \ref{sec:product} of the Appendix. This is used for the more robust protocols described in Sections \ref{sec:box} and \ref{sec:robust}.

\section{Creating an entangled black hole}
\label{sec:entangledbh}

Now we address the question of creating the entangled black hole. In particular, we want to implement step (iii) of the protocol of Equation \eqref{eq:protocol} where the quantum memory $\ket{b}_{N_{B}}\ket{h}_{N_{H}}$ is copied onto ordinary matter and collapsed into a black hole with state $W_{BH}\ket{bh}_{BH}$.
The important ingredient we needed was that the state of the black hole is maximally mixed on its full Hilbert space, so we require that the black hole has maximum entropy for a given mass. This can be done
by starting with a black hole of some arbitrarily small mass $M_0$, and then adding mass to the black hole which is maximally entropic for the amount of energy added. In particular, we can reverse the mining process 
 of Unruh and Wald\cite{unruh1982acceleration,unruh1983mine}. 
Namely, we take a box of thermal radiation, which by definition has maximum entropy for fixed energy.   We then slowly lower the box down to the black hole. We want to add the entropy of the box's contents to the black hole in such a way that we increase the black hole mass by the smallest possible amount. Classically, the increase of black hole mass could be made to be zero, since we could lower the box to the black hole horizon, and at this point it has redshifted away all of its mass as seen at infinity. However, taking into account the buoyancy force that the box feels due to the thermal state of the vacuum, we can only lower the box to its floating point just above the horizon and it is here that it has minimal energy. The box is then opened, delivering the box's entropy to the black hole with minimal mass increase to the black hole. 
During this process, the black hole will be emitting Hawking radiation, and the question is whether we can grow the black hole faster than it's evaporating. Indeed, this is the case.
The mining process has been studied recently, via a process of lowering many ropes to the horizon of the black hole~\cite{brown2012tensile}. This allows radiation to more easily escape the angular momentum potential barrier of the black hole. Removing the
angular momentum barrier means that the radiation can flow out, and flow in
without reflecting. That lack of reflection also means that you can feed  radiation into the
black hole down the ropes as fast as you want, since the radiation
does not reflect back out. The rate of black hole evaporation in such a case, is still proportional to $M^3$ as it was before, but with a smaller proportionality constant. In the reverse process, the bottom end of the ropes are placed on the horizon of the black hole, while the top end of the ropes are placed in thermal contact with thermal radiation at a temperature larger than the black hole temperature. Doing so will grow the black hole, as it is just the reverse of the situation considered in \cite{brown2012tensile} -- the mass transferred to the black hole is done so at a rate just larger that its rate of evaporation.

At the end of this process, there will be a black hole system $BH$ and the radiation $E$ which escaped during this process. Since the radiation that escapes, is via the process described in Equation \eqref{eq:page-map}, the radiation $E$ will be in a product state with $BH$ and further, when system $B$ is ejected into the zone, it will be product with $BE$ provided that the interactions between $E$ and $B$ are small, which we can take them to be, as we can take $B$ to be the radiation which has just left the black hole, while $E$ was emitted earlier, during the black hole's creation. We have thus created the situation required for Equation \eqref{eq:productBE} to hold. In Section \ref{sec:robust}, we describe a more robust protocol, where all we require is that the mutual information $I(B:HE):=S(B)+S(HE)-S(BHE)$ between $B$ and $HE$ is not too large.

\subsection{Creating an entangled black hole in a box}
\label{ss:creatingbox}

We now turn to at the creation of memory entangled with system composed of a maximally entropic black hole enclosed in a box~\cite{hawking1976black,hawking1983thermodynamics} and in thermal equilibrium with the radiation in the box. This situation is needed for our second protocol, described in Section \ref{sec:box}. 
The important ingredient we need for that protocol is that the entire system $BHE$ is in the maximally mixed state, where $E$ is the escaped radiation trapped in the box. In the previous example, because the Hilbert space outside the black hole is far larger than the original Hilbert space of the black hole, $E$ is not maximally mixed on the full Hilbert space. However, by enclosing the black hole in a box which traps the radiation $E$, we can make our entire system sit in a maximally mixed subspace. As a result,
the subsystems $HE$ of the black hole and any escaped radiation is close to being in a product state with the zone photons $B$, satisfying Equation \eqref{eq:productBE}. This is true because any unitary on a maximally mixed state, keeps the state maximally mixed, and any part of a maximally mixed state is in a product state with every other part. Thus, Equation \eqref{eq:productBE} will hold, regardless of the types of interactions between $B$, $H$, and $E$, so that even if black hole evaporation is not of the type described by Equation \eqref{eq:page-map}, the second protocol of Section \ref{sec:box} will still work. In particular, one might imagine alternative models of evaporation, which could entangle degrees of freedom of the zone with those of the black hole, taking an arbitrary state of the black hole, to an arbitrary state of the black hole and zone i.e. $\ket{h}\ket{b}\rightarrow\ket{\psi_{bh}}$. We would thus like our gedanken experiment to work for any such dynamics.

\begin{figure}
\includegraphics[width=10cm]{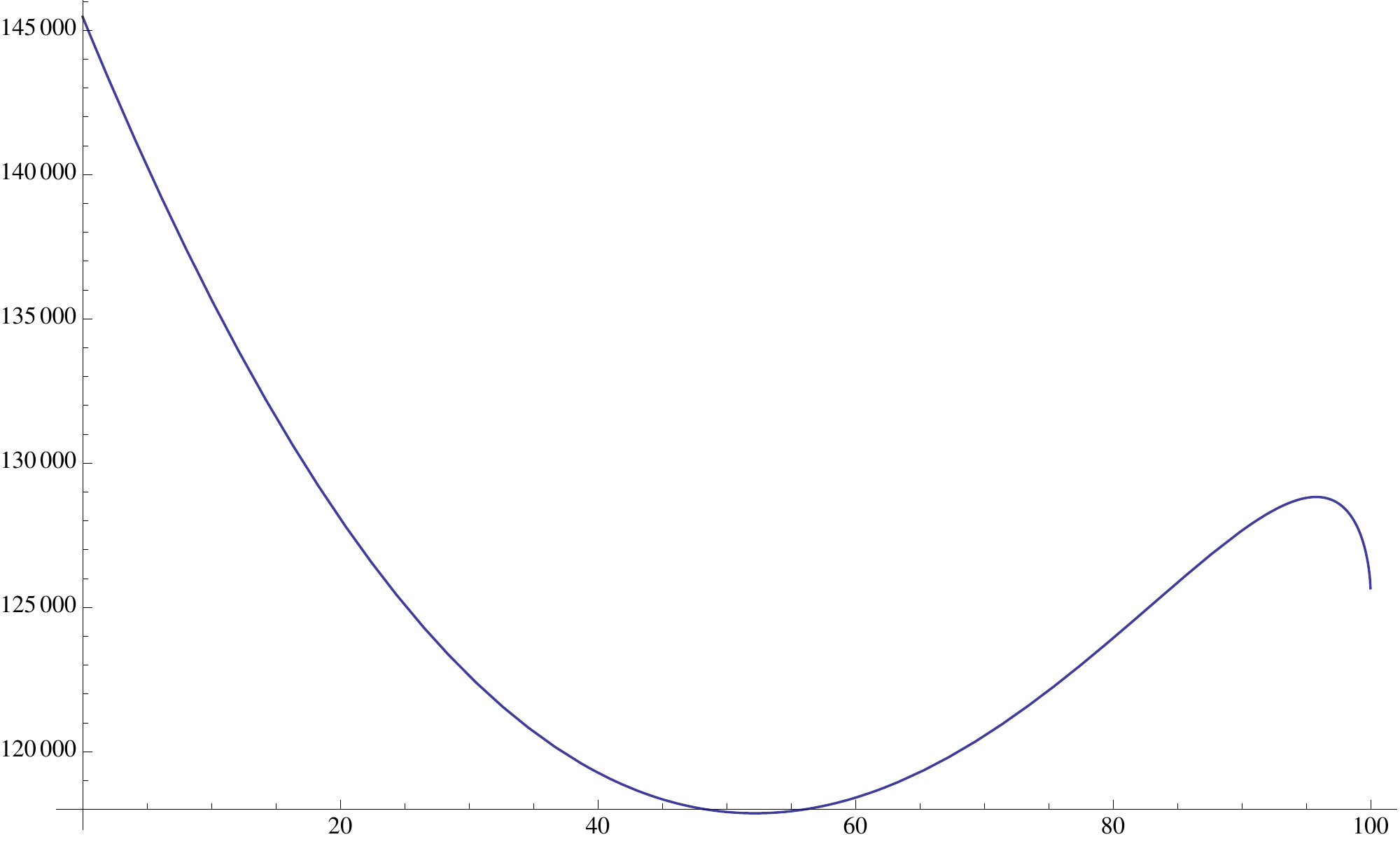}
	\caption{The total entropy $S$ of a black hole in a box of thermal radiation as a function of black hole mass $M$. The entropy $S_{rad}$ of thermal radiation in a box of volume $V$ and temperature $T_{rad}$ is $\alpha V T_{rad}^3$ with $\alpha$ a constant which depends on the species of particles. The entropy $S_{bh}$ of a black hole of mass $M$ in the interior of the box is $4\pi M^2$. Fixing the total energy $E=M+E_{rad}$, we find the total entropy 
	$S=S_{rad}+S_{bh}$ plotted above has one minimum and one maximum when $T_{rad}=T_{bh}$. These correspond to the roots of the equation 
	$(\frac{3}{4\alpha})^4EM^4-M^5-\frac{V}{(8\pi)^4}=0$). The system is stable when the entropy is a maximum, which occurs when $M>4E/5$ \cite{hawking1976black} and unstable at the minimum $M=M_0$ when $M<4E/5$.  We can create a black hole with mass $M>M_0$, using the reverse mining protocol from the previous section, trapping any escaped radiation in the box. Then the black hole will just grow by itself until the total system reaches maximum entropy.
	}
\label{fig:bhbox}
\end{figure}

We thus want to create a state of the black hole and box, which is maximally mixed. I.e at fixed energy, it is a mixture of all possible configurations. The state of such a system, with high probability, will be the state with maximum entropy -- a black hole in thermal equilibrium with its own radiation. We start with a box of radiation and total energy $E$ which is maximally entangled with a quantum memory, so the state of the radiation is maximally mixed over all possible degrees of freedom. We then create a small seed black hole out of a pure state (e.g. by collapsing a shell of dust) inside the box, and then slowly and adiabatically feed the thermal radiation into the black hole, via the reverse of the Unruh-Wald process, slowly growing the black hole. All the while, the thermal radiation inside the box remains in equilibrium with the growing black hole.  Figure \ref{fig:bhbox} describes the process in more detail.

Following this process, the black hole and box will not be in a completely mixed state, since one has started with a seed black hole of mass $M_0$ which was in a pure state. However, we can make it maximally mixed, by performing one of a set of random unitaries on the radiation inside the box, with the unitary chosen conditional on a memory. I.e. we perform 
$\sum_i\proj{i}\otimes U^i_{BE}$ with $\proj{i}$ acting on some additional memory system $M_U$.  We know from the result of \cite{GroismanPW04}, that as long as the number of such unitaries $U^i_{BE}$ is larger than the mutual information $I(H:BE)$, then it will make the state $\rho_{BHE}\approx\rho_{BE}\otimes
\rho_H$ with $\rho_{BE}$ maximally mixed. If this is done a few times while the evaporation dynamics causing mixing between $H$ and $BE$, then this will  succeed in bring the state of the systems inside the box, into a state which is maximally mixed on the subspace spanned by all states inside the box of fixed total energy. As a result, when particles evaporate into the zone $B$,  Equation \eqref{eq:productB} will be satisfied, regardless of the model of black hole evaporation.

\section{An alternative experiment using a memory entangled with a black hole in a box}
\label{sec:box}

We now give a second protocol, which has the advantage that it works for black hole evaporation models which are more exotic that that of Equation \eqref{eq:page-map}.
At its core, we make use of a  property of maximally entangled states -- namely, that instead of acting on one half of a maximally entangled state, one can equivalently act on the other half. For example, one can see that for the $d$-dimensional maximally entangled state $\ket{\psi^+_d}_{MB}=\sum_k\ket{kk}_{MB}/\sqrt{d}$
\begin{align}
\id_{M}\otimes U_B\ket{\psi^+_d}_{MB}= U_{M}^T\otimes\id_{B}\ket{\psi^+_d}_{MB}
\label{eq:uustar}
\end{align}
since,
\begin{align}
\sum_{jk}\bra{jj}U_M^*U_B\ket{kk}/{d}&=\sum_{jk}\bra{k}U_M^\dagger\ket{j}\bra{j}U_B\ket{k}/{d} \\
&=1
\end{align}
Instead of acting $U$ on one share of this maximally entangled state, one can equivalently act $U^T$ on the other share, or equivalently,  $\ket{\psi^+_d}_{MB}$ is invariant
under the action of $U_M^*\otimes U_B$.

 
We now discuss a second protocol which requires creating a black hole in a box~\cite{hawking1976black,hawking1983thermodynamics}.  For this protocol,
imagine then that we have 
created a black hole in a box, and in thermal equilibrium with its own radiation, and that all degrees of freedom inside the box are maximally 
entangled with a quantum memory $M$ composed of three quantum registers $M_H$, $M_B$ and $M_E$ as described in Section \ref{ss:creatingbox}.
\begin{align}
\ket{\Psi(0)}=\frac{1}{\sqrt{|B||H||E|}}
\sum_{h,b}\ket{b}_{M_B}\ket{h}_{M_H}\ket{e}_{M_E} W_{BHE} \ket{bhe}_{BHE}
\end{align}
where once again, we take $W_{BHE}\ket{bhe}_{BHE}$ to initially be the state of the black hole with $W_{BHE}$ being the unitary of the formation process of the black hole and box.  $H$ is take to be the degrees of freedom associated with the black hole, $B$ the photons which are in the zone, and $E$ additional degrees of freedom associated
with the thermal state inside the box. 
Note that due to Equation \eqref{eq:uustar}, Alice can act $W^*_{BHE}$ on her quantum memory in order to  produce the state
\begin{align}
\ket{\Psi'}=\frac{1}{\sqrt{|B|}}
\sum_{h,b,e}\ket{b}_{M_B}\ket{b}_B\otimes \frac{1}{\sqrt{|H|}}\ket{h}_{M_H} \ket{h}_{H}\otimes\frac{1}{\sqrt{|E|}}\ket{h}_{M_E} \ket{h}_{E}
\label{eq:decodeduustar}
\end{align}
in which each of the qubits of her quantum memory is entangled with a single degree of freedom identified with the black hole, the zone, or the remaining thermal radiation.
Now, we let the black hole evaporate further, which we can take to be a unitary evolution $U_{BHE}$ which takes 
degrees of freedom which are inside the black hole and associated with system $H$ to degrees of freedom in the zone $B$ (and visa versa, since the total system is in equilibrium). 
The degrees of freedom on $E$ are also coupled to the zone photons $B$, and since nothing hinges on it, we write this process as a global unitary acting on the entire system inside the box, taking  $\ket{bhe}_{BHE}$ to $\ket{\psi_{bhe}}_{BHE}$. We thus have a total state of 
\begin{align}
\ket{\Psi(t)}=\frac{1}{\sqrt{|B||H||E|}}
\sum_{h,b,e}\ket{b}_{M_B}\ket{h}_{M_H}\ket{e}_{M_E} V_{BHE}W_{BHE} \ket{bhe}_{BHE}
\label{eq:before}
\end{align}

Now, Alice wants to decode a degree of freedom which was inside the black hole, and is now in the zone, in order to produce pure state entanglement between zone photons and
qubits of her memory. Starting from the state of Equation \eqref{eq:before}, she wants to formally act $W^\dagger V^\dagger$ on the $BHE$ system. However, using the relation
from Equation \eqref{eq:uustar}, she can instead do this by acting the transpose $V^*W^*$ to her memory to go from the state of Equation \eqref{eq:before} to the state of Equation \eqref{eq:decodeduustar}. Once again, since she can perform  $V^*W^*$ on the quantum memory, she can do this even before the black hole is created, and so once again, by precomputing $V^*W^*$ on the memory, the black hole will emit perfectly correlated entangled pairs into the zone as with the first protocol of Section \ref{sec:alternative}.

\section{A more robust experiment: Extracting mutual information rather than entanglement}
\label{sec:robust}

There is another way to make this experiment more robust against possible correlations between the zone photons and the black hole degrees of freedom, and it has the advantage of not requiring the creation of a black hole in thermal equilibrium with photons in the box. In the original AMPS experiment, one attempts to witness a violation of  monogamy of entanglement, by decoding pure state entanglement between the zone photons, and the early time Hawking radiation. This contradicts the fact that the zone photons are also in a pure entangled state with the infalling Hawking radiation. However, if the zone photons $B$ are maximally entangled with the infalling Hawking photons ${\overline B}$, then not only can they not be entangled with another system, they can't even be classically correlated with another system. This can be expressed in a variety of forms\cite{koashi2004monogamy}, and we can use a particularly simple one here. If the state on $B{\hat B}$ is pure (and hence $S(B{\hat B})=0$), then
\begin{align}
I(M:B)&\leq I(M:B\hat{B})\\
&=S(M)-S(MB\hat{B})\\
&=0
\end{align}
where the first line is just strong subadditivity, and the rest just follows from the fact that a pure state must be product with any other state.

As a result, we do not need to decode pure entanglement, but just demonstrate that correlations exist between the zone photons, and the quantum memory. Thus, it's still possible to witness a violation of monogamy, even if the zone photons are correlated with the black hole. All we need is that they are also correlated with the memory.
We now demonstrate a protocol to do this. It has the advantage that since we do not require the black hole to be in equilibrium with its radiation, we 
do not need the box, and thus we adapt the protocol of Section \ref{sec:alternative}.
The setup will be slightly different, because we will not require that the particles $B$ which escape into the zone, have the same dimension Hilbert space as the memory $M_B$. We thus represent the state of the memory entangled with the black hole and escaped radiation as
\begin{align}
\ket{\Psi(0)}=\frac{1}{\sqrt{|B||H||E|}}
\sum_{h,b}\ket{b}_{M_B}\ket{h}_{M_H}\ket{e}_{M_E} W_{C\rightarrow BHE} \ket{bhe}_{C}
\end{align}
where the unitary $W_{C\rightarrow BHE}$ represents the black hole formation process, followed by the evaporation process where the part $B$ is evaporated into the zone.
It's inevitable that some photons, even a lot, will escape the black during the formation process, so we  
denote this  system of escaped photons by $E$, and what remains inside the black hole by $H$. 

Let us first show that there is mutual information between the zone photons and the quantum memory. If the dynamics $W_{C\rightarrow BHE}$  of the black hole are sufficiently  scrambling\footnote{Dynamics are considered to be sufficiently scrambling if they randomise a quantum state, as described for example, in \cite{randomization}. A unitary chosen from the uniform distribution, is scrambling with high probability, which tends to $1$ as the dimension of the input grows. A unitary chosen from a set of 2-designs\cite{dankert2006exact}, will also be sufficiently scrambling with high probability.}, 
and if enough photons leave the black hole, then the memory $M_B$ will be close to a product state with $HE$. This occurs provided the system $B$ is large enough\cite{FQSW,HaydenPreskill,dupuis2010one}. In the case where the black hole dynamics consist of a typical unitary followed by ejection of system's into the zones\cite{bhlock,HaydenPreskill,sekino2008fast}, this occurs provided $B$ is slightly larger than $M_B$. More generally, how large $B$ must be, will depend on the model of black hole evaporation, and we examine this in more detail in the Appendix. However, we just require that the total dynamics is sufficiently scrambling, and the fact that some photons $E$ have previously escaped  is not a concern since we may just regard them as being part of system $H$ for our purposes.

Let us assume such dynamics, and the fact that $S(M_B)\approx S(B)$, with $B$ just slightly larger than $M_B$. In such a case,
the state of the black hole and memory $M_B$ will become close to product form\cite{FQSW,HaydenPreskill,dupuis2010one}, and so, since purifications are unique up to a unitary on the purifying system, the state of the entire system is, up to a unitary $U$ on $M_{HE}B$ given by
\begin{align}
\ket{\psi}_{HEM_{\overline{HE}}}\otimes\ket{\psi}_{M_B\overline{B}}
\label{eq:decoupled}
\end{align}
with $M_{\overline{HE}}$ and $\overline{B}$ the output systems of $U$. Note that here, we are just imagining that the unitary $U$ is applied for the purposes of calculating the mutual information between $M_B$ and $B$. Equation \eqref{eq:decoupled} implies that $M_B$ is purified by the remainder of $M$ and $B$ and so we have
 that $S(MB)=S(M)-S(M_B)$. 
Assuming the evaporation models from \cite{bhlock,HaydenPreskill,sekino2008fast} (see the Appendix for general models), and since $S(M_B)\approx S(B)$ we have
$-S(B|M):=S(M)-S(MB)\approx S(B)$.
The quantity $-S(B|M)$ is known as the coherent information\cite{schumacher1996quantum} and it is a lower bound on the amount of one way distillable entanglement between the zone $B$ and the memory $M$, if only classical communication is allowed from the zone to the individual who has the memory\cite{devetak2005distillation}. Since each bit of entanglement gives two bits of mutual information, this means that there exists at least $I(B:M)=2S(B)$ bits of mutual information between the zone and the memory $M$.

Now, although there is substantial mutual information between $B$ and $M$, the memory $M$ is large, and so, to verify that these correlations would take many measurements and thus a long time. So 
Alice will compress her mutual information onto a small memory register as follows. Imagine that someone in the zone, distills entanglement, by performing a measurement on $B$ and communicating the result $i$ of the measurement to Alice. Alice could then
perform a unitary $U_i$ on her quantum memory, which would result in pure maximal entanglement between $B$ and a register $M_{\overline{B}}$. However, $U_i$ is a computationally hard unitary to perform, so she will have to implement it before the black hole has been created. Thus, Alice should guess which $i$ she will receive, and perform $U_i$ on her memory. In general, the number of possible measurement outcomes $i$ is $2^{I(B:HE)}$\cite{devetak2005distillation}. Again, if the evaporation is of the  \cite{bhlock,HaydenPreskill,sekino2008fast}  type, then $I(B:HE)$ will be zero or very small. However, we may imagine more exotic dynamics such that it's non-zero. If Alice guesses $i$ correctly, then she succeeds in distilling $ 2S(B)$ bits of mutual information, and if she guesses wrongly, she might distill no mutual information. The amount of mutual information she succeeds is thus given by $I(B:M)2^{-I(B:HE)}$. In order to verify this, she can dive into the zone with her memory $M_{\overline{B}}$ and verify the correlations in $BM_{\overline{B}}$ (this would require many runs of the experiment). Alternatively, there is a 
probability of $2^{-I(B:HE)}$ that Alice will witness a violation of non-monogomy. While one might want to always verify  a violation of the laws of quantum mechanics 
in order to conclude that
something is amiss with our present theories, a non-vanishing probability of witnessing such a violation, should lead us to the same conclusion.
 
\section{Superluminal signalling near black holes}
\label{sec:chsh}

\begin{figure}
\includegraphics[width=10cm]{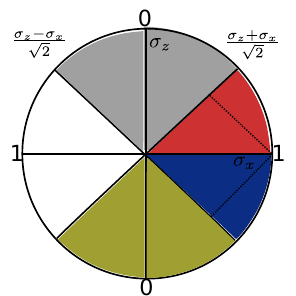}
	\caption{If Bob wishes to communicate a $1$ he measures in the $\sigma_x$ basis. In the example above, we imagine he does so, and gets the result up (the point on the circle where the blue and red wedge meet). Alice now measures $M$ in the $(\sigma_z+\sigma_x)/\sqrt{2}$ basis, and ${\hat B}$ in the $(\sigma_z-\sigma_x)/\sqrt{2}$. Here, she has $P_{win}=1/2+1/2\sqrt{2}$ of getting up in the first measurement (the projection onto the axis where the grey and red region meet), as well as the same probability of getting down on the second measurement (the projection where the green and blue region meet). This gives here a probability of $3/4$ of correctly guessing the basis of Bob's measurement, and hence the bit he is trying to communicate.
	}
\label{fig:chsh}
\end{figure}

To emphasis that a non-zero probability of witnessing a violation of non-monogomy is enough to challenge the idea that black holes preserve information (or posit that firewalls exist), we note here that this would imply a non-vanishing probability of being able to send signals faster than light. This point is also worth making, because it will allow us to impose restrictions on modifications of quantum mechanics such as those examined in \cite{muller2012black}, where it was found that in some generalisations of quantum theory, information from a black hole can appear to be thermal throughout the black hole evaporation process.
We now exhibit a protocol to signal faster than light using our modified AMPS experiment. The idea is to note that after decoding entanglement between her memory and the zone, Alice could jump into the black hole, so that she holds both  the quantum memory $M_B$ and some of the infalling Hawking photons ${\hat B}$. Then, an observer Bob in the zone $B$ could attempt to violate a Bell inequality, both with Alice via $M$ and via ${\hat B}$. Since Bob is maximally entangled both with $M_B$ and with  ${\hat B}$, it will be possible to violate a Bell inequality with both systems.  However, it has been noted that the no-signalling criteria imposes restrictions on being able to violate the CHSH inequality\cite{clauser1969proposed}  with two parties\cite{toner2006monogamy}. A protocol to signal faster than light, is as follows: If Bob wants to signal a $1$, he measures in the Pauli $\sigma_x$ basis, and if he wants to communicate $0$, he measures in the $\sigma_z$ basis. Alice now measures $M$ in the $(\sigma_z+\sigma_x)/\sqrt{2}$ basis, and ${\hat B}$ in the $(\sigma_z-\sigma_x)/\sqrt{2}$ (see Figure \ref{fig:chsh}). The probability that the measurement on $M$ correctly determines whether the outcome of Bob's measurement was up or down is $P_{win}=1/2+1/2\sqrt{2}$, while the measurement on $B$ has the same probability of correctly determining whether the measurement outcome was one of two cases, with case 1 being up in the $\sigma_z$ basis or down in the $\sigma_x$ basis (depending on what basis Bob measured in), and case 2 being the opposite -- down in the $\sigma_z$ basis or up in the $\sigma_x$. Combing these two results gives a probability of correctly guessing Bob's measurement basis, and hence the communicated bit, of $1-2P_{win}+2P^2_{win}=3/4$. By performing this protocol on many entangled pairs, they can make the probability of successfully communicating, arbitrarily close to one.

\bibliography{../common/refgrav2,../common/refjono,../common/refjono2}

\providecommand{\href}[2]{#2}\begingroup\raggedright\begin{thebibliography}{10}

\bibitem{almheiri2013black}
A.~Almheiri, D.~Marolf, J.~Polchinski, and J.~Sully, {\it Black holes:
  complementarity or firewalls?},  {\em Journal of High Energy Physics} {\bf
  2013} (2013), no.~2 1--20.

\bibitem{braunstein2009entangled}
S.~L. Braunstein and K.~{\.Z}yczkowski, {\it Entangled black holes as ciphers
  of hidden information},  {\em arXiv preprint arXiv:0907.1190} (2009).

\bibitem{hawking-bhinfoloss}
S.~W. Hawking, {\it Breakdown of predictability in gravitational collapse},
  {\em Phys. Rev. D} {\bf 14} (Nov, 1976) 2460--2473.

\bibitem{hawking-unpredictability}
S.~W. {Hawking}, {\it {The unpredictability of quantum gravity}},  {\em
  Communications in Mathematical Physics} {\bf 87} (Dec., 1982) 395--415.

\bibitem{preskill-infoloss-note}
For background reading, we refer the reader to the review of John Preskill,
  hep-th/9209058.

\bibitem{coffman2000distributed}
V.~Coffman, J.~Kundu, and W.~K. Wootters, {\it Distributed entanglement},  {\em
  Physical Review A} {\bf 61} (2000), no.~5 052306.

\bibitem{bennet-monogomy}
The term was initially coined by Charlie Bennett, Tata Institute for
  Fundamental Research, Mumbai (1999).

\bibitem{koashi2004monogamy}
M.~Koashi and A.~Winter, {\it Monogamy of quantum entanglement and other
  correlations},  {\em Physical Review A} {\bf 69} (2004), no.~2 022309.

\bibitem{tHooft-bhcompl}
G.~{t'Hooft}, {\it On the quantum structure of a black hole},  {\em Nuclear
  Physics B} {\bf 256} (1985), no.~0 727 -- 745.

\bibitem{tHooft-bhcompl-string}
G.~t'Hooft, {\it The black hole interpretation of string theory},  {\em Nuclear
  Physics B} {\bf 335} (1990), no.~1 138 -- 154.

\bibitem{susskind-bhcompl}
L.~Susskind, L.~Thorlacius, and J.~Uglum, {\it The stretched horizon and black
  hole complementarity},  {\em Phys. Rev. D} {\bf 48} (Oct, 1993) 3743--3761.

\bibitem{bps}
T.~{Banks}, M.~E. {Peskin}, and L.~{Susskind}, {\it {Difficulties for the
  evolution of pure states into mixed states}},  {\em Nuclear Physics B} {\bf
  244} (Sept., 1984) 125--134.

\bibitem{unruh-wald-onbps}
W.~G. Unruh and R.~M. Wald, {\it Evolution laws taking pure states to mixed
  states in quantum field theory},  {\em Phys. Rev. D} {\bf 52} (Aug, 1995)
  2176--2182.

\bibitem{OR-intrinsic}
J.~Oppenheim and B.~Reznik, {\it Fundamental destruction of information and
  conservation laws},  {\em Arxiv preprint arXiv:0902.2361} (2009).

\bibitem{unruh2012decoherence}
W.~Unruh, {\it Decoherence without dissipation},  {\em Philosophical
  Transactions of the Royal Society A: Mathematical, Physical and Engineering
  Sciences} {\bf 370} (2012), no.~1975 4454--4459.

\bibitem{page-unitary-evap}
D.~Page, {\it Is black-hole evaporation unitary},  {\em Phys. Rev. Lett.} {\bf
  44} (1980) 301.

\bibitem{marolf2013gauge}
D.~Marolf and J.~Polchinski, {\it Gauge-gravity duality and the black hole
  interior},  {\em Physical review letters} {\bf 111} (2013), no.~17 171301.

\bibitem{giddings2012nonviolent}
S.~B. Giddings, {\it Nonviolent nonlocality},  {\em arXiv preprint
  arXiv:1211.7070} (2012).

\bibitem{susskind2012singularities}
L.~Susskind, {\it Singularities, firewalls, and complementarity},  {\em arXiv
  preprint arXiv:1208.3445} (2012).

\bibitem{papadodimas2012infalling}
K.~Papadodimas and S.~Raju, {\it An infalling observer in ads/cft},  {\em arXiv
  preprint arXiv:1211.6767} (2012).

\bibitem{bousso2013complementarity}
R.~Bousso, {\it Complementarity is not enough},  {\em Physical Review D} {\bf
  87} (2013), no.~12 124023.

\bibitem{jacobson2012boundary}
T.~Jacobson, {\it Boundary unitarity without firewalls},  {\em arXiv preprint
  arXiv:1212.6944} (2012).

\bibitem{banks2013no}
T.~Banks and W.~Fischler, {\it No firewalls in holographic space-time or matrix
  theory},  {\em arXiv preprint arXiv:1305.3923} (2013).

\bibitem{almheiri2013apologia}
A.~Almheiri, D.~Marolf, J.~Polchinski, D.~Stanford, and J.~Sully, {\it An
  apologia for firewalls},  {\em arXiv preprint arXiv:1304.6483} (2013).

\bibitem{maldacena2013cool}
J.~Maldacena and L.~Susskind, {\it Cool horizons for entangled black holes},
  {\em arXiv preprint arXiv:1306.0533} (2013).

\bibitem{shenker2013black}
S.~H. Shenker and D.~Stanford, {\it Black holes and the butterfly effect},
  {\em arXiv preprint arXiv:1306.0622} (2013).

\bibitem{mathur2013flaw}
S.~D. Mathur and D.~Turton, {\it The flaw in the firewall argument},  {\em
  arXiv preprint arXiv:1306.5488} (2013).

\bibitem{van2013evaporating}
M.~Van~Raamsdonk, {\it Evaporating firewalls},  {\em arXiv preprint
  arXiv:1307.1796} (2013).

\bibitem{larjo2013black}
K.~Larjo, D.~A. Lowe, and L.~Thorlacius, {\it Black holes without firewalls},
  {\em Physical Review D} {\bf 87} (2013), no.~10 104018.

\bibitem{lloyd2013unitarity}
S.~Lloyd and J.~Preskill, {\it Unitarity of black hole evaporation in
  final-state projection models},  {\em arXiv preprint arXiv:1308.4209} (2013).

\bibitem{harlow2013quantum}
D.~Harlow and P.~Hayden, {\it Quantum computation vs. firewalls},  {\em arXiv
  preprint arXiv:1301.4504} (2013).

\bibitem{susskind2013black}
L.~Susskind, {\it Black hole complementarity and the harlow-hayden conjecture},
   {\em arXiv preprint arXiv:1301.4505} (2013).

\bibitem{unruh-newton}
Bill Unruh, "Bohr, Penrose, and Hawking", presented at the Isaac Newton
  Institute, workshop on quantum gravity and quantum information, Cambridge,
  Dec. 14th, 2004
  http://www.newton.cam.ac.uk/webseminars/pg+ws/2004/qisw05/1214/unruh/.

\bibitem{bhlock}
J.~A. {Smolin} and J.~{Oppenheim}, {\it {Locking Information in Black Holes}},
  {\em Physical Review Letters} {\bf 96} (Feb., 2006) 081302,
  [\href{http://xxx.lanl.gov/abs/hep-th/0507287}{{\tt hep-th/0507287}}].

\bibitem{HaydenPreskill}
P.~Hayden and J.~Preskill, {\it Black holes as mirrors: quantum information in
  random subsystems},  {\em J. High Energy Phys.} {\bf 09} (2007), no.~120.

\bibitem{sekino2008fast}
Y.~Sekino and L.~Susskind, {\it Fast scramblers},  {\em Journal of High Energy
  Physics} {\bf 2008} (2008) 065.

\bibitem{giddings2013quantum}
S.~B. Giddings and Y.~Shi, {\it Quantum information transfer and models for
  black hole mechanics},  {\em Physical Review D} {\bf 87} (2013), no.~6
  064031.

\bibitem{how-merge}
M.~Horodecki, J.~Oppenheim, and A.~Winter, {\it Partial quantum information},
  {\em Nature} {\bf 436} (2005) 673--676,
  [\href{http://xxx.lanl.gov/abs/quant-ph/0505062}{{\tt quant-ph/0505062}}].

\bibitem{how-merge2}
M.~Horodecki, J.~Oppenheim, and A.~Winter, {\it Quantum state merging and
  negative information},  {\em Comm. Math. Phys.} {\bf 269} (2006) 107,
  [\href{http://xxx.lanl.gov/abs/quant-ph/0512247}{{\tt quant-ph/0512247}}].

\bibitem{uhlmann1976transition}
A.~Uhlmann, {\it The" transition probability" in the state space of
  a*-algebra},  {\em Reports on Mathematical Physics} {\bf 9} (1976), no.~2
  273--279.

\bibitem{unruh1982acceleration}
W.~G. Unruh and R.~M. Wald, {\it Acceleration radiation and the generalized
  second law of thermodynamics},  {\em Physical Review D} {\bf 25} (1982),
  no.~4 942.

\bibitem{unruh1983mine}
W.~G. Unruh and R.~M. Wald, {\it How to mine energy from a black hole},  {\em
  General Relativity and Gravitation} {\bf 15} (1983), no.~3 195--199.

\bibitem{brown2012tensile}
A.~R. Brown, {\it Tensile strength and the mining of black holes},  {\em arXiv
  preprint arXiv:1207.3342} (2012).

\bibitem{hawking1976black}
S.~W. Hawking, {\it Black holes and thermodynamics},  {\em Physical Review D}
  {\bf 13} (1976), no.~2 191.

\bibitem{hawking1983thermodynamics}
S.~W. Hawking and D.~N. Page, {\it Thermodynamics of black holes in anti-de
  sitter space},  {\em Communications in Mathematical Physics} {\bf 87} (1983),
  no.~4 577--588.

\bibitem{GroismanPW04}
B.~Groisman, S.~Popescu, and A.~Winter, {\it Quantum, classical, and total
  amount of correlations in a quantum state},  {\em Physical Review A} {\bf 72}
  (2005), no.~3 032317.

\bibitem{randomization}
P.~Hayden, D.~Leung, P.~Shor, and A.~Winter, {\it Randomizing quantum states:
  Constructions and applications},  {\em Commun. Math. Phys.} {\bf 250(2)}
  (2004) 371--391.

\bibitem{dankert2006exact}
C.~Dankert, R.~Cleve, J.~Emerson, and E.~Livine, ``Exact and approximate
  unitary 2-designs: constructions and applications.'' quant-ph/0606161.

\bibitem{FQSW}
A.~{Abeyesinghe}, I.~{Devetak}, P.~{Hayden}, and A.~{Winter}, {\it {The mother
  of all protocols: Restructuring quantum information's family tree}},
  \href{http://xxx.lanl.gov/abs/quant-ph/0606225}{{\tt quant-ph/0606225}}.

\bibitem{dupuis2010one}
F.~Dupuis, M.~Berta, J.~Wullschleger, and R.~Renner, {\it One-shot decoupling},
   {\em arXiv preprint arXiv:1012.6044} (2010).

\bibitem{schumacher1996quantum}
B.~Schumacher and M.~A. Nielsen, {\it Quantum data processing and error
  correction},  {\em arXiv preprint quant-ph/9604022} (1996).

\bibitem{devetak2005distillation}
I.~Devetak and A.~Winter, {\it Distillation of secret key and entanglement from
  quantum states},  {\em Proceedings of the Royal Society A: Mathematical,
  Physical and Engineering Science} {\bf 461} (2005), no.~2053 207--235.

\bibitem{muller2012black}
M.~P. M{\"u}ller, J.~Oppenheim, and O.~C. Dahlsten, {\it The black hole
  information problem beyond quantum theory},  {\em Journal of High Energy
  Physics} {\bf 2012} (2012), no.~9 1--32.

\bibitem{clauser1969proposed}
J.~F. Clauser, M.~A. Horne, A.~Shimony, and R.~A. Holt, {\it Proposed
  experiment to test local hidden-variable theories},  {\em Physical Review
  Letters} {\bf 23} (1969) 880--884.

\bibitem{toner2006monogamy}
B.~Toner, F.~Verstraete, D.~Gross, K.~Audenaert, J.~Eisert,
  H.~Brusheim-Johansson, J.~Hansson, J.~Hodges, P.~Cappellaro, T.~Havel,
  et~al., {\it {Monogamy of Bell correlations and Tsirelson's bound}},  {\em
  Arxiv preprint quant-ph/0611001} (2006) 67.

\bibitem{wolf2008area}
M.~M. Wolf, F.~Verstraete, M.~B. Hastings, and J.~I. Cirac, {\it Area laws in
  quantum systems: mutual information and correlations},  {\em Physical review
  letters} {\bf 100} (2008), no.~7 070502.

\end{thebibliography}\endgroup

\appendix

\section{A black hole in thermal equilibrium is in a product state with its own radiation}
\label{sec:product}

Consider a black hole $H$ which is in a thermal state, and in equilibrium with its own thermal radiation (which we take to be photons in the zone $B$, but the result will be general).
We denote their joint thermal state by $\rho_{BH}$ Then, following \cite{wolf2008area} we note that the thermal state is the state which minimises the free energy $F_{BH}=E(BH)-TS(BH)$, with $E(BH)$ the total energy and $T$ the temperature. In particular, we have $F(\rho_{BH})\leq F(\rho_B\otimes\rho_H)$, with $\rho_H=\tr_B\rho_{BH}$ and similarly for $\rho_B$. We thus have
\begin{align}
E(\rho_{BH})-TS(BH)&\leq E(\rho_B\otimes\rho_H)-T[S(B)+S(H)]\nonumber\\
I(B:H)&\leq\beta [E(\rho_B\otimes\rho_H)-E(\rho_{BH})]\nonumber\\
&\leq 0
\end{align}
with the last line following from the fact, that there is a horizon separating $H$ from $B$, and thus, the energy of $B$ cannot be a function of the correlations between $B$ and $H$ as it would allow the experimenter at $B$ to determine properties of $H$ beyond its total energy.

If on the other hand, we start with a black hole of energy $E$, and let it evaporate into an empty box with walls just outside the zone, and don't condition on the mass of the black hole $M$, then the total state of the black hole and zone will not necessarily be in a product state. That's because the system will evolve to a state in which the photons in the zone $B$ look thermal (and hence maximally entropic), and the entropy of the black hole will be maximally entropic since the evaporation is adiabatic, but the total state of the black hole $H$ plus zone $B$ does not have maximally entropy. This can be seen from the fact that for fixed total energy $E$, the number of states which have $M=E$ and the energy of the zone photons $E_B=0$ is small compared with the maximum number of states possible if we just fix the total energy $M+E_B=E$ as we do for the micro-conical ensemble. Thus, creating a black hole and letting it evaporate into an empty box, will produce a state which has locally maximally entropy, but the global entropy will not be maximal, thus there will be mutual information $I(H:B)$ between the zone and black hole as $I(H:B)=S(H) + S(B)-S(BH)$ and $S(BH)$ cannot be maximal if the evolution is unitary. 

\section{Decoupling for general black hole evaporation models}
\label{sec:general}

Here, we examine under what conditions, the black hole radiation is in a product state with the black hole when the dynamics of black hole evaporation is different to 
that considered by Page in Figure \ref{fig:page}, i.e. a model where the black hole undergoes a unitary evolution, followed by ejection of a system out of the horizon.
In such a case, the evaporated photons in $B$ may be correlated with the interior of the black hole even for maximally entangled black holes. To verify the
violation of monogamy of entanglement, we would then need to implement the protocol from Section \ref{sec:robust}, and so we are interested in knowing 
when the black hole $H$ is no longer correlated with $M_B$. We base our analysis on the single-shot decoupling theorem, proven in \cite{dupuis2010one}.

Let us imagine that initially, $H$ and $B$ are inside the black hole, undergoing unitary dynamics $U$ which we take to be sufficiently scrambling. Finally, system $B$ is evaporated in some way, into the zone. Here, we take the dynamics of this evaporation process to be an arbitrary completely positive trace preserving (CPTP) map ${\cal T}_B$, which takes the subsystem $B$ inside the black hole into the zone.  We will describe the CPTP map by how it looks from the point of view of the black hole interior i.e. ejection of the system $B$ to the zone, corresponds to tracing out subsystem $B$, since from the point of view of the interior, the subsystem $B$ is now gone. More generally, we have some arbitrary channel ${\cal T}_B$ which acts
on $B$, with the output of the map being what remains inside the black hole. We will describe this map by the Choi-Jamio{\l}kowski state $\tau_{B{\hat B}}$ which is
the (possibly sub normalised) state produced when ${\cal T}_B$ acts on the maximally entangled state on $B{\hat B}$. 
We can thus also view ${\cal T}_B$ as a map from $B\rightarrow{\hat B}$.
We now define the quantum collision entropy
\begin{align}
H_2(X|Y)_{\rho}:=\overset{\sup}{\sigma_Y} -\log\tr\left[ \left((\id_X\otimes\sigma_Y^{-1/4})\rho_{XY}(\id_X\otimes\sigma_Y^{-1/4})\right)^2\right]
\end{align}
where the supremum is taken over subnormalised states $\sigma_Y$.

We know apply the single shot decoupling theorem~\cite{dupuis2010one}, saying that for almost all $U$ (i.e. $U$ chosen according to the Haar measure), and any $\mu>0$,
\begin{align}
||{\cal T}_B(U\rho_{BHM_B}U^\dagger)-\tau_{{\hat B}}\otimes\frac{\id_H}{|H|}||
&\leq
\frac{1}{\mu} 2^{
-\frac{1}{2}H_2(B|M_B)_\rho-\frac{1}{2}H_{2}(B|{\hat B})_\tau
}\\
&=
\frac{1}{\mu} 2^{\frac{1}{2}\log|M_B|
- \frac{1}{2}H_{2}(B|{\hat B})_\tau
}
\end{align}
where in the last line, we've made use of the fact that the initial state on $BM_B$ was maximally entangled, and in the first line, the terms involving $H$ have cancelled.
We thus require that the dynamics are such that $H_{min}(B|{\hat B})_\tau>\log|M_B|+c$ with $c$ determining how close the black hole will be decoupled from $M_B$.
We can verify that if  ${\cal T}_B$ is the tracing out operation, then $H_{min}(B|{\hat B})_\tau=\log|B|$, and so we just require that $B$ be larger than $M_B$ by a few bits.

 Because we are considering the case when $\rho_{BH}$ is maximally mixed, $\tau_{B{\hat B}}$ also determines how much correlation remains between the black hole and the zone photons, which in this case, is just $I(B:{\hat B})_\tau$. The protocol
 of Section \ref{sec:robust} thus succeeds in signalling superluminally, with a probability of $2^{I(B:{\hat B})_\tau}$. It is worth noting that if the black hole is to remain
 maximally entropic, as we expect since the evaporation process is adiabatic. then the state of $\hat B$ should be maximally mixed.  However, we expect it to be maximally
 mixed on a system of very small dimension, since the evaporation process does need to lower the entropy of the black hole to zero at a rate proportional to the number of photons
 which are emitted.


\setcounter{secnumdepth}{3} 
{\bf Acknowledgements} 
JO thanks Daniel Harlow, Patrick Hayden, Douglas Stanford and John Preskill for helpful discussions, and the Royal Society for support.

\bibliographystyle{JHEP}

\end{document}